\documentclass[10pt,conference]{IEEEtran}
\usepackage[T1]{fontenc}
\usepackage{amsmath}
\usepackage{booktabs}
\usepackage{array}
\usepackage[hidelinks]{hyperref}
\hypersetup{pdfauthor={Mohamad Mohebifar; Alex Bit},pdftitle={OneCodeReviewBench: Typed Decisions for Code Review and Search},pdfsubject={Typed decisions for code review and search}}

\title{OneCodeReviewBench: Typed Decisions\\for Code Review and Search}
\author{\IEEEauthorblockN{Mohamad Mohebifar and Alex Bit}
\IEEEauthorblockA{Codemod Inc.}}

\begin{document}
\maketitle

\begin{abstract}
Code review requires finding functional bugs, memory-safety errors, security weaknesses, and performance problems across languages and software stacks. Tools can express bounded parts of this work as decisions about supplied code. We present OneCodeReviewBench, 1,968 synthetic cases across 12 programming languages and YAML, with Boolean, categorical, and ordered answers. Explicit contracts specify runtime assumptions, source scope, and answer criteria; separate evidence records support reference review. One author reviewed 150 cases; two additional engineers reviewed 100 of those cases. We compare ten model--adapter systems on the same cases and contracts: OpenAI Decisions reaches 81.3\% Boolean balanced accuracy; Kev-27B reaches 76.8\% Choice accuracy and Score mean absolute error of 0.367. Excluding two families with incomplete contracts preserves these leaders. Additional comparisons cover 586 cases with raw symbol search and 40 with a bounded Python import resolver; the resolver matches all 40 references within its supported profile. A versioned artifact reproduces these analyses and retains the 85 cases with matching engineer answers; the reported review subset excludes two incomplete-contract cases. Remaining reviewer disagreements and repaired inputs require resolution; repository-scale transfer and matched latency/cost evaluation remain future work.
\end{abstract}


\section{Introduction}
\label{sec:introduction}
Code-review and search tools may call a generative model for every file. A decision model with a bounded answer could filter candidate files or answer a narrow question directly. Adaptive cascades in FrugalGPT~\cite{chen2024frugalgpt} and stronger/weaker-model routing in RouteLLM~\cite{ong2025routellm} study related quality--cost tradeoffs. Here \emph{decision model} describes an answer interface, not a particular architecture. We evaluate these components for review and search; measuring their end-to-end speed and cost requires separate experiments.

Code review examines functional correctness, memory and resource safety, potential security vulnerabilities, concurrency, and performance across languages and stacks. Bounded questions include whether an authorization rule holds, whether a scheduling pattern can deadlock, or which resource-lifetime outcome follows from supplied inputs. Search questions locate API uses that warrant inspection. These decisions need explicit file scope, runtime assumptions, and answer criteria: finding an API call alone does not establish a bug. Structural tasks require static-tool comparators; work on lexical retrieval also motivates inexpensive code-search baselines~\cite{zhang2021bagofwords}.

Reference quality is an evaluation concern regardless of whether people or models construct the cases. Label errors can affect benchmark comparisons~\cite{northcutt2021labelerrors}, and PrimeVul documents labeling and duplication issues in vulnerability datasets~\cite{ding2024vulnerability}. We therefore record how references were produced, which cases engineers reviewed, what mechanical checks establish, and which reviews remain unresolved. AI generation describes the construction method; it does not by itself determine whether an answer is right or wrong.

We ask: \emph{How reliably do model--adapter systems answer bounded review and search questions under explicit code-task contracts?} We contribute a multilingual task collection and paired evaluation protocol that link each decision to its supplied files, assumptions, answer rubric, and reference-review evidence. Researchers can compare interfaces on shared cases and trace ambiguous answers to specific contract rules. We demonstrate these uses through ten-system results, engineer review, static checking, and repairs prompted by a contract audit. The contribution lies in the dataset and its review records; the answer types and input/answer-key separation are established techniques.

\subsection{Related work and task scope}
CodeReviewer studies real changes and reviews~\cite{li2022codereviewer}; CRUXEval evaluates short-program input/output prediction~\cite{gu2024cruxeval}; CodeMMLU includes categorical code questions~\cite{manh2025codemmlu}; CodeSense constructs semantic references from real-project execution traces~\cite{roy2025codesense}. Table~\ref{tab:positioning} contrasts their units with ours. CodeSearchNet evaluates natural-language retrieval with expert relevance judgments~\cite{husain2019codesearchnet}; our inventory tasks specify a closed file universe and matching/counting rules.

\begin{table}[t]
\caption{Task units and reference evidence define the comparison scope.}
\label{tab:positioning}
\centering
\begin{tabular}{@{}p{0.24\columnwidth}p{0.69\columnwidth}@{}}
\toprule
Work & Evaluation unit and reference basis \\
\midrule
CodeReviewer & Real changes and review activities \\
CRUXEval & Short-function input/output reasoning; executable relationship \\
CodeMMLU & Categorical questions about code understanding \\
CodeSense & Real-project semantic reasoning; execution traces \\
This work & Declared files and decision rules; synthetic references, engineer review, and checking records \\
\bottomrule
\end{tabular}
\end{table}

Execution-backed evaluation provides another reference-checking path~\cite{liu2023is}. LOTUS optimizes semantic operators with accuracy guarantees relative to reference algorithms~\cite{patel2025lotus}. Our collection supplies review-specific decisions and closed-file inventory contracts in a common evaluation format, with evidence linked to each reference. These units complement change-level review, program execution, and open-corpus retrieval benchmarks. Practical transfer requires separate evaluation on repository-grounded tasks.

\section{Case Format and Decision Contracts}
\label{sec:protocol}
A case contains evaluator-visible files, context, a question, and answer criteria. The dataset reference is stored separately for offline scoring, and review/checking records link to case identifiers and input hashes. In the schema, \texttt{state} and \texttt{question} supply evaluator input; the historical \texttt{gold} field stores the generated reference. This separation supports evaluation and subsequent reference review without exposing the explicit answer key to a model request.

Boolean criteria define complementary true/false conditions. Choice defines mutually exclusive categories within the stated domain. All 560 Score cases define four ordered levels, 0--3, with task-specific criteria. For source counts, these represent zero, one, two, and at least three matching sites. Other rubrics cap nesting depth or simultaneously held resources at three; a Python isolation-policy rubric instead assigns higher levels to fewer violations. Score MAE is measured in the units of each stated rubric. A count rubric is not a severity scale, and increasing levels do not always mean increasing risk. A search contract states the file universe, production/test roles, target identity, counting unit, and treatment of nested/unreachable code. A behavior contract states the entry point, inputs, runtime, and observation boundary.

\subsection{Worked source-inventory case}
Case \texttt{cd942e58ec34661e26ac} counts direct calls to \texttt{asyncio.shield} in production source. Its manifest declares both files production and specifies static inspection under standard-library import semantics:
\begin{verbatim}
# module_a.py
import asyncio as aio
async def hold(task):
    return await aio.shield(task)

# module_b.py
import asyncio
async def delay():
    return await asyncio.sleep(0)
\end{verbatim}
We compress blank lines here. The archived target call is at \texttt{module\_a.py}:5, column 17. The generated reference and import-resolving checker both yield level 1. Evidence records the call-node location and imported identity; a string, local same-named function, or excluded test-file call would not count. This is an API inventory that can guide inspection, not a cancellation-safety judgment.

Behavior questions instead require reasoning about an outcome. For example, a Tokio case asks whether two tasks can complete on a single-thread executor: one holds a synchronous mutex across an await, while another blocks that executor acquiring the mutex before sending the wake-up signal. The scheduling and lock semantics determine the answer. Memory-lifetime, authorization, and database-access cases similarly state the conditions needed to assess a potential defect. Such bounded fixtures isolate review decisions; they do not establish whole-repository coverage.

\subsection{Review, abstention, and repair}
Engineers answer cases from the supplied contracts and record supporting evidence. A mechanical checker derives an answer before comparing it with the stored reference and abstains on unsupported semantics. Reviewer uncertainty about a language differs from a case missing a decisive rule. When we repair a contract, we retain the historical input and require fresh review and inference on the new version, so old predictions do not stand in for results on repaired inputs.

\section{Collection and Reference Review}
\label{sec:collection}
We constructed 500 draft templates: 246 Boolean, 127 Choice, and 127 Score. Forty were individually authored; expansion added behavior/configuration scenarios and search targets across 23 ecosystem groups covering 12 programming languages and YAML. Agent-assisted template review addressed answer overlap and scope.

GPT-6-Sol generated fixtures, context, and initial reference answers for a profile selecting 164 templates and requesting 2,000 cases. Sixteen of 656 synthesis jobs exhausted after upstream HTTP 500 responses, leaving 1,968 cases from 161 templates. Table~\ref{tab:coverage} distinguishes dataset coverage, engineer review, and abstract syntax tree (AST) checking; the latter two are complementary and can overlap.

\begin{table}[t]
\caption{Dataset and review coverage. AST counts show the shield checker; the 40-case import comparison in Section~\ref{sec:evaluation} overlaps nine of these cases. Engineer/static coverage can overlap. The 24 adapter controls are outside this collection.}
\label{tab:coverage}
\centering
\begin{tabular}{@{}lrrrr@{}}
\toprule
Type & Cases & Templates & Reviewed & AST \\
\midrule
Boolean & 728 & 92 & 50 & 0 \\
Choice & 680 & 34 & 50 & 0 \\
Score & 560 & 35 & 50 & 15 \\
\midrule
Total & 1,968 & 161 & 150 & 15 \\
\bottomrule
\end{tabular}
\end{table}

The median fixture contains two files and 27 source lines; the 90th percentiles are four files and 65 lines. Source-inventory questions account for 840 cases: 440 Choice and 400 Score. Of these, 808 come from expanded inventory templates (440 Choice and 368 Score); 32 Score cases come from two original templates, 16 each. The remaining 1,128 cases include behavior, configuration, and structural decisions. Small fixtures make case-level reasoning inspectable; they leave repository-scale context retrieval and coordination outside the present evaluation.

\subsection{Engineer review of 150 cases}
Experienced software engineers reviewed the 150-case packet. One author supplied all 150 responses, and two additional, paid engineers reviewed disjoint 50-case batches. They made the final decisions themselves, with possible AI assistance for questions; case-level assistance was not logged. The packet excludes reference answers and system predictions.

Of 100 cases with two responses, 85 answers agree, seven differ, and eight second reviewers abstained. Several abstention notes identify language/framework unfamiliarity. Fifty cases still need a second response, and disagreements, abstentions, and known contract repairs remain in the follow-up queue. These records provide human-review evidence; we have not converted unresolved responses into a final adjudicated key.

The full-packet reviewer's answers match 149 of 150 stored references. The sole mismatch concerns a nested-body exclusion missing from the packet; a second case omits category priority despite a matching answer. A subsequent source audit finds no visible answer contradiction in the other 148 cases, without executing all pinned frameworks. We record these outcomes separately from agreement between reviewers. The quota sample and unresolved cases do not support a population correctness estimate.

The packet balances answer types and Boolean reference labels, and balances search/nonsearch cases within Choice and Score while spreading selection across templates. It therefore supports case review rather than a simple pooled estimate of label error across the entire run. A population estimate would require sampling weights and completed resolution.

\section{Model Comparison and Case-Quality Checks}
\label{sec:evaluation}
We evaluate whether the common contracts support paired model comparisons, and how review and checking records identify case-quality issues. We report the main results alongside easy controls and device/option-order diagnostics to distinguish typed-output validity from task agreement.

\subsection{Ten systems on identical cases}
Every system returned a valid prediction for the same 1,968 identifiers. Table~\ref{tab:systems} introduces their access modes and implementations. Jev used \texttt{jev-1.13.0}; OpenAI Decisions used \texttt{gpt-6-luna}. Azure used deployment \texttt{decision-1}, reporting \texttt{microsoft-decision-1}; its October 23, 2026 run preserves raw requests, responses, and UTC timestamps. Laya uses SDK 0.3.6 and revision \texttt{1c5edc17}; Von uses SDK 1.1.1 and revision \texttt{d8bb5e07}. Kev-0.8B/4B/9B use v1.0 checkpoints; Kev-27B uses v2. The artifact records checkpoint/source revisions.

\begin{table}[t]
\caption{Systems and access modes. The four Kev checkpoints are separate rows in the results. Hosted backend architectures are not isolated by this comparison.}
\label{tab:systems}
\centering
\begin{tabular}{@{}>{\raggedright\arraybackslash}p{0.31\columnwidth}>{\raggedright\arraybackslash}p{0.62\columnwidth}@{}}
\toprule
System & Access and implementation \\
\midrule
Jev~\cite{typesafe2026jev} & TypeSafe hosted typed-decision API \\
OpenAI Decisions~\cite{openaiDecisions} & Hosted predicate, choice, and score API \\
Azure Decision-1~\cite{microsoft2026decision} & Hosted native Microsoft decision API \\
OpenJev (SemIf)~\cite{semifSource} & Local direct option scorer using Qwen3.5-4B \\
Laya~\cite{layaModelCard} & Local typed-decision SDK; \texttt{convaiinnovations/laya}, \texttt{typed-decisions} subfolder \\
Von~\cite{vonSource} & Local typed-decision SDK; \texttt{von-1.1.0} checkpoint \\
Kev (four sizes)~\cite{kevSource} & Local Qwen-based decision checkpoints via upstream Kev API and MLX \\
\bottomrule
\end{tabular}
\end{table}

For Jev, OpenAI, Azure, Laya, Von, and Kev, adapters retain the returned Boolean probability, selected category, and Score scalar. Rounded probability vectors are normalized where required by the local SDK adapters; the reported Score scalar is retained. SemIf instead scores explicit answer options: the adapter takes the true-option probability for Boolean, the highest-probability category for Choice, and $\sum_{i=0}^{3} i p_i$ for Score. This conversion does not establish probability calibration. Adapters serialize the same files, context, and criteria differently; valid outputs do not establish identical effective tokens or absence of truncation. Some historical timestamps and Azure's pinned backend revision are unavailable. These are model--adapter comparisons, not isolated architectures or controlled scaling.

Table~\ref{tab:models} reports balanced Boolean accuracy, Choice accuracy, and Score mean absolute error (MAE) against dataset references. Boolean uses a 0.5 threshold and averages accuracy within each reference class; Choice uses exact category matching; Score uses the supplied scalar on the 0--3 rubric. Thus the measures quantify reference agreement, with human-review coverage described in Section~\ref{sec:collection}. We retain all original reference labels for this historical run while resolving review items separately.

\begin{table}[t]
\caption{Paired results on 728 Boolean, 680 Choice, and 560 Score cases. BA is balanced accuracy (\%); Choice is exact-match accuracy (\%); Score MAE is lower-is-better. Baselines are always-true Boolean, uniform Choice, and the in-sample level-2 Score constant. Reference provenance and review coverage appear in Table~\ref{tab:coverage}.}
\label{tab:models}
\centering
\begin{tabular}{@{}lrrr@{}}
\toprule
System & Bool.\ BA & Choice & Score MAE \\
\midrule
Simple baselines & 50.0 & 24.7 & 0.855 \\
\midrule
OpenAI Decisions & 81.3 & 68.2 & 0.516 \\
Azure Decision-1 & 64.1 & 63.4 & 0.430 \\
Jev & 71.4 & 71.6 & 0.403 \\
OpenJev (SemIf) & 52.6 & 42.9 & 0.727 \\
Laya & 49.4 & 27.8 & 0.839 \\
Von & 50.9 & 26.6 & 0.892 \\
Kev-0.8B & 50.2 & 32.9 & 0.899 \\
Kev-4B & 54.6 & 47.9 & 0.723 \\
Kev-9B & 56.5 & 55.9 & 0.571 \\
Kev-27B & 79.7 & 76.8 & 0.367 \\
\bottomrule
\end{tabular}
\end{table}

OpenAI Decisions has the highest observed Boolean BA; Kev-27B has the highest Choice accuracy and lowest Score MAE. The leading system therefore depends on the decision type. Exploratory paired template-cluster percentile bootstraps use 4,000 draws (seed 20260924). Kev-27B exceeds Jev on Choice by 5.15 points (95\% interval: 1.91--8.53); OpenAI Decisions versus Kev-27B on Boolean and Kev-27B versus Jev on Score have intervals including zero. Kev-27B exceeds Azure on Choice by 13.38 points (8.97--17.79); all six Azure--Kev/OpenAI intervals are in the artifact. Intervals condition on stored references, omit label uncertainty, and are unadjusted for multiple comparisons or shared template lineage. The runs do not measure a matched latency/cost advantage.

Of the 85 cases with matching engineer answers, two belong to the incomplete-contract families. Excluding them leaves 83 cases (28 Boolean, 27 Choice, 28 Score): Kev-27B reaches 89.0\% Boolean BA, 81.5\% Choice accuracy, and Score MAE 0.228; Jev reaches 74.1\%, 77.8\%, and 0.239; Azure reaches 72.8\%, 55.6\%, and 0.301. These scores use stored references supported by the recorded reviewer comparison. Pending cases are excluded. This agreement-selected quota subset is descriptive, not a population sample or final adjudicated key.

\subsection{Easy controls and adapter diagnostics}
We used 24 deliberately easy, AST-checked Python controls outside the collection: eight per answer type, covering literal markers, categorical decisions, and capped counts. Table~\ref{tab:controls} reports the nine systems with saved control results. All returned valid typed outputs, but Von and Kev-0.8B answered only four of eight Boolean controls correctly; Laya and Von also had substantial Score error on these simple cases. SemIf lacks this control run because the available rerun environment did not meet its MPS requirement. These small controls diagnose basic behavior; they neither prove an adapter defect nor set a numerical bound on benchmark performance.

\begin{table}[t]
\caption{Separate easy-control results: eight cases per type. Boolean and Choice report correct answers out of eight; Score reports MAE on levels 0--3. Every row has 24 valid typed outputs. SemIf was not evaluated on this control set.}
\label{tab:controls}
\centering
\begin{tabular}{@{}lrrr@{}}
\toprule
System & Bool. & Choice & Score MAE \\
\midrule
OpenAI Decisions & 8/8 & 8/8 & 0.00000 \\
Azure Decision-1 & 8/8 & 8/8 & 0.00017 \\
Jev & 8/8 & 8/8 & 0.00000 \\
Laya (CPU) & 8/8 & 8/8 & 0.81790 \\
Von (CPU) & 4/8 & 8/8 & 0.93500 \\
Kev-0.8B & 4/8 & 8/8 & 0.48571 \\
Kev-4B & 8/8 & 8/8 & 0.12871 \\
Kev-9B & 8/8 & 8/8 & 0.05169 \\
Kev-27B & 8/8 & 8/8 & 0.00126 \\
\bottomrule
\end{tabular}
\end{table}

Paired CPU/MPS checks for Laya and Von preserved all Boolean, Choice, and modal Score decisions across the 1,968 generated cases and 24 controls. Probability/scalar differences were at most approximately $10^{-4}$. These checks establish agreement only for the paired runs tested. In three Von Choice-order checks (seeds 17, 42, 97), shuffling criteria insertion order while preserving labels and descriptions changed 357, 336, and 350 of 680 generated answers (49.4--52.5\%). The eight Choice controls remained correct under each shuffle. Option order therefore affects this model--adapter evaluation; shuffled seeds are not independent benchmark samples.

\subsection{Source-search baselines}
We define comparator recipes from public query rules before joining reference labels or model predictions. Raw symbol search counts case-sensitive occurrences of the target's terminal identifier, including imports, declarations, comments, and strings; it uses explicit file roles but does not resolve bindings or types. Of 808 expanded inventory cases, 220 designated-root questions are outside its profile, and two lack a role for a supplied configuration file. It answers 586 cases across 34 templates and all 12 languages plus YAML. Table~\ref{tab:search-baselines} compares systems on exactly those answered IDs.

A Python AST comparator resolves direct calls through module-level imports, includes nested/unreachable source, and applies the stated site/file count or role rubric. It excludes instance/type and cross-file resolution, and abstains on ambiguous bindings, escapes, or mutation. It answers 40 of 120 eligible cases across five templates; all match the references. Alias, shadowing, and manifest counterexamples test the supported rules. The two comparator subsets differ, and neither provides a general static-analysis or correctness guarantee. Their results show why review/search evaluation needs both inexpensive source-tool baselines and explicit coverage.

\begin{table}[t]
\caption{Source-search comparisons on each baseline's answered subset. Choice is accuracy (\%); Score is MAE. These are development subsets; the two blocks have different case populations. All ten systems' paired results and abstentions are in the artifact.}
\label{tab:search-baselines}
\centering
\begin{tabular}{@{}lrr@{}}
\toprule
System & Choice & Score MAE \\
\midrule
\multicolumn{3}{@{}l}{Raw-symbol subset: 220 Choice, 366 Score} \\
Raw symbol search & 34.5 & 0.839 \\
OpenAI Decisions & 66.8 & 0.574 \\
Azure Decision-1 & 65.5 & 0.422 \\
Jev & 73.6 & 0.400 \\
Kev-27B & 81.4 & 0.358 \\
\midrule
\multicolumn{3}{@{}l}{Python-import subset: 7 Choice, 33 Score} \\
Import resolver & 100.0 & 0.000 \\
OpenAI Decisions & 42.9 & 0.641 \\
Azure Decision-1 & 57.1 & 0.528 \\
Jev & 57.1 & 0.308 \\
Kev-27B & 42.9 & 0.410 \\
\bottomrule
\end{tabular}
\end{table}

\subsection{Mechanical checking and contract audit}
We checked all 16 generated cases from one \texttt{asyncio.shield} source-count template, selected for bounded static semantics rather than observed model scores. The checker resolves imports/aliases, applies file roles, counts direct call nodes including nested/unreachable sites, and caps the result at three. Fifteen derived answers match the references; one case returns an imported callable outside a direct call and triggers abstention. The procedure records input hashes and file/line evidence. It used Python 3.14.4 with the 3.12 grammar setting, not execution under the declared CPython 3.12.7 runtime. Its scope is this template, not a general Python analyzer.

A source audit identified two cases with omitted rules: exclusion of nested function/class bodies in a Python count, and category priority for co-occurring Go violations. Auditing their families found these omissions across 16 and 20 inputs. We archived 36 repaired inputs separately. The omissions establish that the contracts were incomplete; they do not establish that all 36 answers were wrong. The repaired inputs need fresh responses and inference. Reviewing individual cases thus led us to inspect and repair both template families.

Excluding both families leaves 1,932 cases and preserves the leaders: Kev-27B reaches 76.5\% Choice accuracy and Score MAE 0.366, versus Jev's 71.2\% and 0.400 and Azure's 63.8\% and 0.432; Boolean results are unchanged. Excluding only the two documented cases also preserves the leaders. This sensitivity analysis removes historical inputs; it is not evaluation on repaired versions.

\subsection{Reproduction}
The public reproduction release contains visible inputs, separately keyed references, ten systems' saved predictions, model/source provenance, analysis code, expected outputs, and checksums. Its offline verifier reproduces the metrics, paired intervals, contract exclusions, static slice, source-search baselines, reviewer-agreement subsets, easy controls, and device/option-order diagnostics without model calls. The larger historical archives retain the nine-system snapshot and Azure wire records locally; these are not required by the public verifier. Preserved predictions support reproduction despite endpoint changes. Tokenizer/truncation traces and some historical raw control runs remain unavailable.

\section{Scope and Next Evaluation Steps}
\label{sec:validation}
\textbf{Reference review.}
The 150-case engineer sample and bounded static procedures document complementary evidence; their coverage can overlap. Next steps are resolving the 65-case follow-up queue, expanding review across types/families, and re-reviewing repaired contracts. We preserve disagreements and distinguish expertise abstention from missing assumptions. Neither quota sampling nor profile-specific checking yields a full-run label-error rate.

\textbf{Practical transfer and efficiency.}
The current fixtures are synthetic, and related constructions reduce effective diversity. The generator and one evaluated service are both from OpenAI; provider overlap may affect results, but the present data do not establish its direction or size. Selection for reviewer agreement also favors cases on which reviewers converge. Repository-grounded cases with documented review/search questions are needed to assess transfer. The present source-tool baselines address inventory tasks; behavioral review needs additional task-appropriate analyzers or execution checks. Matched generative/decision-model comparisons should record hardware, batching, latency, and cost at stated correctness levels to test the intended tooling benefit.

\textbf{Release integrity.}
The released artifact includes a dataset card documenting composition, collection, intended uses, provenance, and limitations following datasheet concerns~\cite{gebru2021datasheets}. We release aggregate review counts and subset case identifiers; individual responses, reviewer identities, and private comments are excluded. Participant-consent documentation, the additional engineers' affiliations, and formal ethics-review status were not recorded in the available source materials; no ethics approval or exemption is claimed. Further studies should freeze checked inputs, group templates by lineage, and record effective model inputs.

\section{Artifact Availability and Conclusion}
\label{sec:artifacts}
Version \texttt{20261023-v1} of the dataset and reproduction artifact is publicly archived on Zenodo~\cite{mohebifar2026onecodereviewbench} at \url{https://doi.org/10.5281/zenodo.23604731}. Its offline analyses have been checked from a clean extraction, and uploaded checksums match the verified local files. Codemod Inc. licenses the dataset collection, records, and documentation under CC BY 4.0, and the reproduction software and identified fixture source under Apache-2.0. The release's \texttt{LICENSES.md} defines these material boundaries and applicable exceptions. A separate label-free packet supports new blinded review; the public archive contains references and predictions and must remain separate from those annotation inputs.

OneCodeReviewBench supports paired evaluation of typed code-review and search decisions across ten systems. The release contains 1,968 cases with explicit contracts, engineer review records for 150 cases, source-tool comparisons, and audits that identified missing contract rules. The results distinguish bounded review questions from inventory that deterministic tools can handle within supported profiles. Assessing its use for selecting review/search components requires resolving outstanding reviews, evaluating repository-grounded cases, and measuring efficiency.

\section*{AI Assistance Disclosure}
OpenAI Codex assisted with benchmark and analysis implementation and with manuscript drafting and revision. Anthropic Claude Fable 5.1 provided pre-submission manuscript critique. Dataset generation and possible AI assistance during engineer review are described in Section~\ref{sec:collection}. The human authors retain responsibility for the manuscript and its reported results. The validation items described above remain open.

\IEEEtriggeratref{14}
\bibliographystyle{IEEEtran}

\end{document}